\documentclass[12pt]{article}

\usepackage[english]{babel}

\usepackage{epsfig}
\usepackage{dcolumn}
\usepackage{amsmath}
\usepackage{changebar}
\usepackage{graphicx}
\usepackage{graphicx}
\usepackage{dcolumn} 
\usepackage{bm}      
\usepackage{umlaut}
\begin{document}
\renewcommand{\figurename}{{\bf Fig.}}
\renewcommand{\tablename}{{\bf Tab.}}
\title{\vspace*{-4cm}\hfill {\normalsize\bf BI-TP 2003/21} \\ 
\vspace*{3cm}
The effects of colored quark entropy on the bag pressure}

\author{David E. Miller$^{1,2}$\thanks{om0@psu.edu}
 $\;$ and Abdel-Nasser~Tawfik$^{1}$\thanks{tawfik@physik.uni-bielefeld.de}\\~\\
 {\small $^1$Fakult\"at f\"ur Physik, Universit\"at Bielefeld,
         P.O.Box~100131,} \\
 {\small D-33501~Bielefeld, Germany} \\ 
 {\small $^2$Department of Physics, Pennsylvania State University, 
         Hazleton Campus,} \\
 {\small Pennsylvania 18201, USA} }

\date{}
\maketitle

\begin{abstract}
We study the effects of the ground state entropy of colored quarks upon the
bag pressure at low temperatures. The vacuum expectation values of the quark
and gluon fields are used to express the interactions in QCD ground state
in the limit of low temperatures and chemical potentials. Apparently, the
inclusion of this entropy in the equation of state provides the hadron
constituents with an additional heat which causes a decrease in the
effective latent heat inside the hadronic bag and 
consequently decreases the non-perturbative bag pressure. We
have considered two types of baryonic bags, $\Delta$ and 
$\Omega^-$. In both cases we have found that the bag pressure decreases with
the temperature. On the other hand, when the colored quark ground state
entropy is not considered, the bag pressure as conventionally believed remains
constant for finite temperature.   
\end{abstract}

\noindent
PACS: 12.39.-x \hspace*{2mm} Phenomenological Quark Models, \\
\hspace*{14mm}05.30.-d \hspace*{1.5mm} Quantum Statistical Mechanics

\section{\label{sec:1} Introduction}

The entropy is a concept which has taken on many meanings throughout the
sciences. Its usual sense relates the heat changes to the likelihood of the
related processes at various determined temperatures. In the limit of low
temperatures Planck~\cite{Planck64} noted that a mixture of different
substances retained a finite entropy even at absolute zero. This result 
is quite contrary to the usual interpretation of Nernst's heat theorem, 
for which the entropy should vanish in the low temperature limit. 
It was Schr\"odinger~\cite{Schr} who pointed out a similar observation 
for $N$ atoms each with a two level ground state. 
In this case we should take into consider besides the {\it thermal} entropy
an additional entropy of value $N\ln2$ with Boltzmann
constant $k$ taken to be unity. Schr\"odinger's result~\cite{Schr} can be
readily attained for a system  of $N$ spin one half states. 
Apparently, these results are consistent with the principle of degeneracy
and particle distinguishability. Consequently, it is consistent with the
statistical definition of the quantum entropy using the density 
matrix $\rho$. The latter relates directly to the wavefunction. For a recent
review see~\cite{Geno} and the references therein. This {\it
  quantum} entropy is given by the trace over the quantum states as follows:
\begin{equation}
S = - \hbox{ Tr } \rho \ln \rho \label{eq:0} 
\end{equation}
Although the name "Quantum Entropy" implies
the construction of the density matrix $\rho$ from the quantum states,
the actual mathematical form is well rooted in the laws of classical 
physics~\cite{Planck64}. The entropy of mixing of different types of
ideal gases with constant particle number, volume and temperature, can be
calculated in the same way as Eq.~\ref{eq:0} by replacing 
$\rho$ with $x_i$, which is just the proportion of each constituent 
type $i$ in the total gas system. Thus the additional part of entropy
(entropy of mixing) becomes 
$-{\sum_i}x_i{\ln x_i}$, where the sum has replaced the trace operation.
This expression is clearly a constant independent of the temperature 
so that it must remain at absolute zero~\cite{Planck64}. In the Lie algebra
the quantum fluctuation and correspondingly the quantum entropy are given
by non-zero commuting set. In general, this kind of entropy  represents the
uncertainties in the abundance of information, for which there is an upper
bound of accuracy given by the quantum uncertainty principle~\cite{Deutsch}.\\ 

    In a recent work we have applied these ideas to the quark singlet ground
state of the hadrons~\cite{Mill}. The color symmetry $SU(3)_c$ provides an 
entropy for each of the colored quarks with the value $\ln3$. We have
extended this result to models at finite temperatures~\cite{MiTa}. This
entropy reflects the probability of quark mixing maneuver inside
the hadrons. At vanishing temperature the confined quarks can be viewed as
continuously tousled objects.  In present work we investigate the
contribution of the ground state entropy to the  equation of state for the
colored quarks using the phenomenological bag model for strong
interactions~\cite{DoGoHo}. We  assume that inside the hadron bag all of
the strong interactions at low temperatures $T$ and small quark chemical
potentials $\mu_q$ are included in the quark and gluon condensates. We will
not elaborate here any further on details of the bag model. We apply it as
a simplest analytical model for the QCD equation of state. In this model
and in the low temperature limit the thermodynamical 
quantities of the quarks are given by the non-perturbative bag
pressure, which is widely known as the bag constant ${\mathbf
  B}$. ${\mathbf B}$ gives the gain in the energy density of 
the confined state relative to its value in vacuum
(section~\ref{sec:2}). In particular, we will look at a special model for
baryons, for which the effective degrees of freedom are entirely given by
the quark and gluon colors. All other couplings are taken so that the spin
and flavor are not explicitly considered. 

Hypothetical colored quarks have been suggested as an
intermediate phase in the confinement-deconfinement transition. At high
chemical potentials and low temperatures the hadronic matter has been
conjectured to dissolve into degenerate fermionic quarks. This cold dense
quark matter is believed to exist in the interior of compact stars. The
phenomenological behavior of colored quarks at vanishing $T$ has been
discussed in~\cite{Berg:1986aq}. These are some examples, in which this
part of entropy should be taken into account.   

\section{\label{sec:01}Quark and gluon condensates}

     In the limit of low temperatures and chemical potentials the
interactions in the QCD ground state are expressed in terms of the vacuum
expectation values of the quark and the gluon fields. The calculation of
these vacuum contributions are gotten from the operator product expansion
using the QCD sum rules~\cite{SVZ,ReRuYa}, which has the local operators of
dimension four yielding the main contributions to the
thermodynamics~\cite{Leut,BoMi,Milli}. The pure gluon vacuum expectation 
value is calculated~\cite{Nari} from the product of the field strength tensors
\hbox{$G^a_{\mu \nu}G_a^{\mu \nu}$} including the non-zero renormalization
group  $\beta(g)$-function~\cite{Leut}. 
\begin{eqnarray}
<G^2>_0 &=& \frac{-\beta(g)}{2g^3}  G^a_{\mu \nu}G_a^{\mu \nu} \label{eqq:4}
\end{eqnarray}
where the repeated indices are summed over their range of values.
From here on the subscript in $<\cdots>_0$ refers to $T=0$.
This gluon condensate can be extracted~\cite{SVZ} 
from the charmonium spectrum to yield a consistently estimated~\cite{DoGoHo} 
value of about $1.95\;$GeV/fm$^3$. 

For the quark condensates we consider two extreme cases for vacuum
expectation values:  
\begin{eqnarray}
m_q<\bar{q}q>_0 &=& m_{lq}<\bar{u}u+\bar{d}u>_0\hspace{4.mm} \hbox{pure
                   light quarks}                \nonumber \\
                &=& m_s<\bar{s}s>_0 \hspace{15.mm} \hbox{pure strange quarks}
\end{eqnarray}
The operator for the pion decay relates the product of light quark 
condensate and the corresponding mass to the fixed value \hbox{$-m_{\pi}^2
f_{\pi}^2$}, $f_{\pi}$ is the pion decay factor. For strange quarks we can
apply a similar relation with $-m_{K}^2f_{K}^2$. In this case $f_{K}$
stands for the kaon decay. 
We use for light quark mass \hbox{$m_{q}= m_u\equiv m_d=6\;$MeV} and strange
quark mass $m_s=150\;$MeV. With these values together with 
$m_{\pi}=138\;$MeV and $m_K=496\;$MeV~\cite{PDG} we find the light
and strange quark condensates are $-42\;$MeV/fm$^3$ and $-273\;$MeV/fm$^3$,
respectively. The averaged vacuum contribution to the fields of  
dimension four in the equation of state can be calculated from the
operator product expansion~\cite{Nari} using 
\begin{eqnarray}
<\Theta_{\mu}^{\mu}>_0 &=& <G^2>_0 + m_q<\bar{q}q>_0 \label{totalconden}
\end{eqnarray}
As in~\cite{Leut} the thermal decay of the quark and gluon
condensates is different. Meanwhile the quark condensate 
shows a thermal dependence up the order of $T^8$, the gluon
condensate has a much slower decay. Therefore, we assumed in
Eq.~\ref{totalconden} that the two condensates 
are $T$-independent, especially in the low temperature limit.
Furthermore, we remark here that both  
condensates have the same color singlet ground state $0^{++}$ often 
associated with the scalar glueball state~\cite{HanJoPet}.\\

   As we mentioned above, we have chosen two extreme cases for  
the baryonic structure. If we look at the spin $3/2$ structure of 
$\Delta$ and $\Omega^-$ as examples of these ground state structures, 
we get a minimal effect from the spin entanglement and flavor mixing. 
Singly, the quark and gluon colors are assumed to be asymmetric. 
Nevertheless, we include the degeneracy factor due to the quark spins 
for the integration over the momenta. The usual sum over the flavors 
is replaced by a factor of three in the baryons. We keep the degeneracy 
factor due to the gluon spins since both polarizations are possible. 
We shall use the trace anomaly for the substitution 
of the above extracted values for the vacuum condensates into the equation 
of state. As discussed above, it is known~\cite{Leut,BoMi,Milli} that the
temperature has very  
little effect on these values at temperatures well below $100\;$MeV. Thus 
we can look at the quantum effects of the entropy on the bag pressure 
${\mathbf B}$ for low temperatures. We do not look at the explicit
dependence of the bag pressure on the quark chemical potential $\mu_q$.
Another reason for excluding the $\mu_q$ dependence is that the 
colored quark entropy is not given in terms of $\mu_q$ (Eq.~\ref{eq:6}).

\section{\label{sec:2}The Equation of State}

At finite temperatures and zero quark chemical potentials we choose the
grand canonical partition function ${\mathcal{Z}}(T,V,\mu_q=0)$ in order to
write down the equation of state in terms of difference between the energy
density and the pressure \hbox{$\varepsilon(T)-3p(T)$}. Hereupon, we 
describe the expectation values for the gluon and quark condensates from
the equation of state as,
\begin{eqnarray}
\lim_{T\rightarrow 0}<\Theta_{\mu}^{\mu}>_T &=&
\varepsilon(T)-3p(T)\label{eqtheta1} 
\end{eqnarray}
where the repeated indices represent the sum over the Lorentz indices.
In the formulation of the bag model~\cite{DoGoHo} the thermodynamics is 
usually included with a bag energy density \hbox{$\varepsilon=+\mathbf B$}
and a bag pressure \hbox{$p=-\mathbf B$}, which generally represents the
energy density and the confining pressure of the bag against the vacuum. 
After applying the first law of thermodynamics which relates the entropy
density to the internal energy and pressure densities 
\begin{eqnarray}
T s(T) &=& \epsilon(T) + p(T)
\end{eqnarray}
we find that Eq.~\ref{eqtheta1} reads
\begin{eqnarray}
\left<\Theta_{\mu}^{\mu}\right>_{T} &=& T \left[\frac{3{\cal
   S}_{q,3}(T)}{V} + s(T)\right] - 4\; [p(T) - {\mathbf B}] 
\label{eq:2}
\end{eqnarray}
\noindent
Here we include the ground state entropy ${\cal S}_{q,3}$~\cite{Mill,MiTa} and 
the bag pressure ${\mathbf B}$. The latter is usually assumed as
independent of the parameters of the ensemble. The l.h.s is estimated as in
Eq.~\ref{totalconden}. For $T\rightarrow0$ we get back Eq.~\ref{eqtheta1},
for which $<\Theta_{\mu}^{\mu}>_{0}\rightarrow 4{\mathbf B}$, according to the
bag model. \\

Since the effective degrees of freedom considered here are merely the
colors of quarks and gluons, we can apply our model~\cite{MiTa} in
calculating ${\cal S}_{q,3}$.  
\begin{eqnarray} 
{\cal S}_{q,3}(T) &=& -\frac{1}{3}\left(1-{e^{-m_q/T}}\right)
  \ln\left[\frac{1}{3}\left(1-{e^{-m_q/T}}\right)\right] \nonumber \\
  &&  -2 z \left[\ln(z)\cos(\theta)-\theta\sin(\theta)\right] \label{eq:6}
\end{eqnarray}
where the values for $z$ and $\theta$  respectively  are 
\begin{eqnarray}
        z &=& {\frac{1}{3}} \left(1 + e^{-m_q/T}~+~e^{-2 \;
        m_q/T} \right)^{1/2} \;, \nonumber \\
  \theta &=& \arctan{\left(\frac{\sqrt{3} \; e^{-m_q/T}}
          {2 + e^{-m_q/T}}\right)}\;.\label{eq:deftheta}
\end{eqnarray}
The other thermodynamic quantities in Eq.~\ref{eq:2} are given as follows:
\begin{eqnarray} 
p(T) &=& \frac{3}{\pi^2}\;T\;\int_0^{\infty} k^2\; dk
   \ln\left(1+e^{-\frac{\epsilon(k)}{T}}\right)
   +\frac{8\pi^2}{45}T^4
   \label{eq:3} \\
s(T)&=& \frac{3}{\pi^2}\;\frac{1}{T}\;\int_0^{\infty} k^2\; dk
  \frac{\epsilon(k)}{e^{\frac{\epsilon(k)}{T}}+1}+  
  \frac{p_q(T)}{T} + \frac{32\pi^2}{45} T^3 \label{eq:5}
\end{eqnarray}
\noindent
where $\varepsilon(k)^2=m_q^2+k^2$ is the single particle energy.
Equations~\ref{eq:3} and~\ref{eq:5} give the pressure and  
the entropy density inside the baryonic bag in depending on
$T$ in the usual way. In Eq.~\ref{eq:5} the expression
$p_q(T)$ means just the quark contribution to the
pressure in the first term of Eq.~\ref{eq:3}. 
The contributions of gluons to these quantities are also taken into
account, for which we considered the spin degeneracy due to the possible
polarization. However, for the sake of simplicity, we have given the gluon 
radiation in the grand canonical partition function in an
approximated form\footnote[1]{At high temperatures the pressure of 
  equilibrated ideal bosonic gas of gluons reads \begin{eqnarray}
p(T)_{gluon} &=& \frac{8}{45} \pi^2\;T^4\; \left[1- \frac{15 \alpha_s}{4
               \pi} + \cdots\right] \nonumber 
\end{eqnarray} where  $\alpha_s$ is the running strong 
coupling constant, which depends on $T$. At low
temperatures we can take
\hbox{$\alpha_s\rightarrow 0$}.} 
which, for instance, appears in the second term of Eq.~\ref{eq:3}. This
approximation is appropriate since $T$ remains small
compared to $\Lambda_{QCD}$.\\
  
By using the values of vacuum condensates given above, we can numerically 
solve Eq.~\ref{eq:2} to get the dependence of the bag pressure 
${\mathbf B}$ upon the temperature $T$. 
The universal constant of the free-space bag pressure is still
debatable. A temperature dependence with the effective Lagrangian for
gauge fields has been reported in~\cite{MR}. Also according to the finite
density QCD sum rules a large reduction in ${\mathbf B}$ is
expected~\cite{FJL}. For the two special baryonic structures of the hadron
bag model, we shall compare the constancy of the bag pressure with and without 
the ground state entropy ${\cal S}_{q,3}$ for the colored quarks. The inclusion
of ground state entropy term~\cite{Mill,MiTa} given in Eq.~\ref{eq:6} when
set into the equation of state Eq.~\ref{eq:2} is to be viewed as providing with
additional heat and therefore has the effect of decreasing the 
thermal pressure of colored quarks and gluons. This leads to decreasing in
the value of ${\mathbf B}$ needed to preserve the hadron bag's stability
against  the force of the outside vacuum.

\section{\label{sec:3} Results and Discussion}

\begin{figure}
\centerline{\includegraphics[width=12.cm]{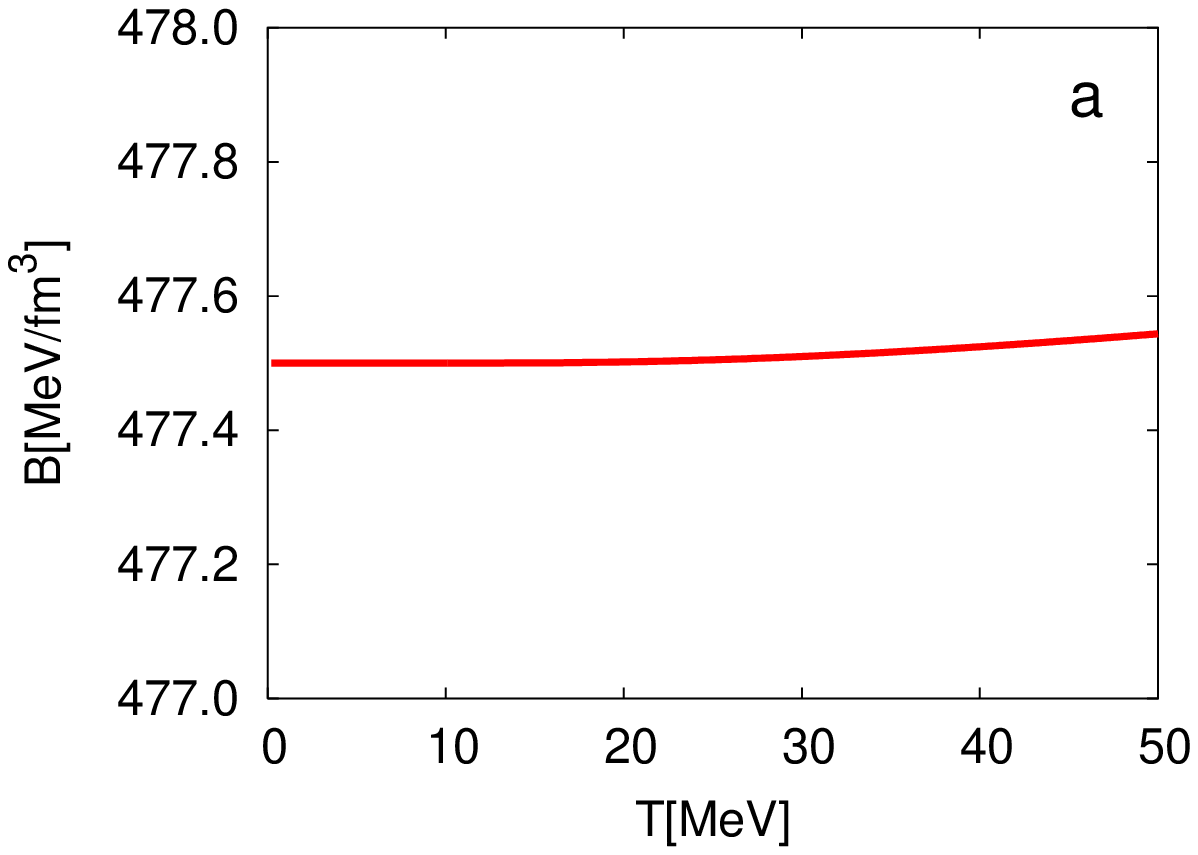}}%
\centerline{\includegraphics[width=12.cm]{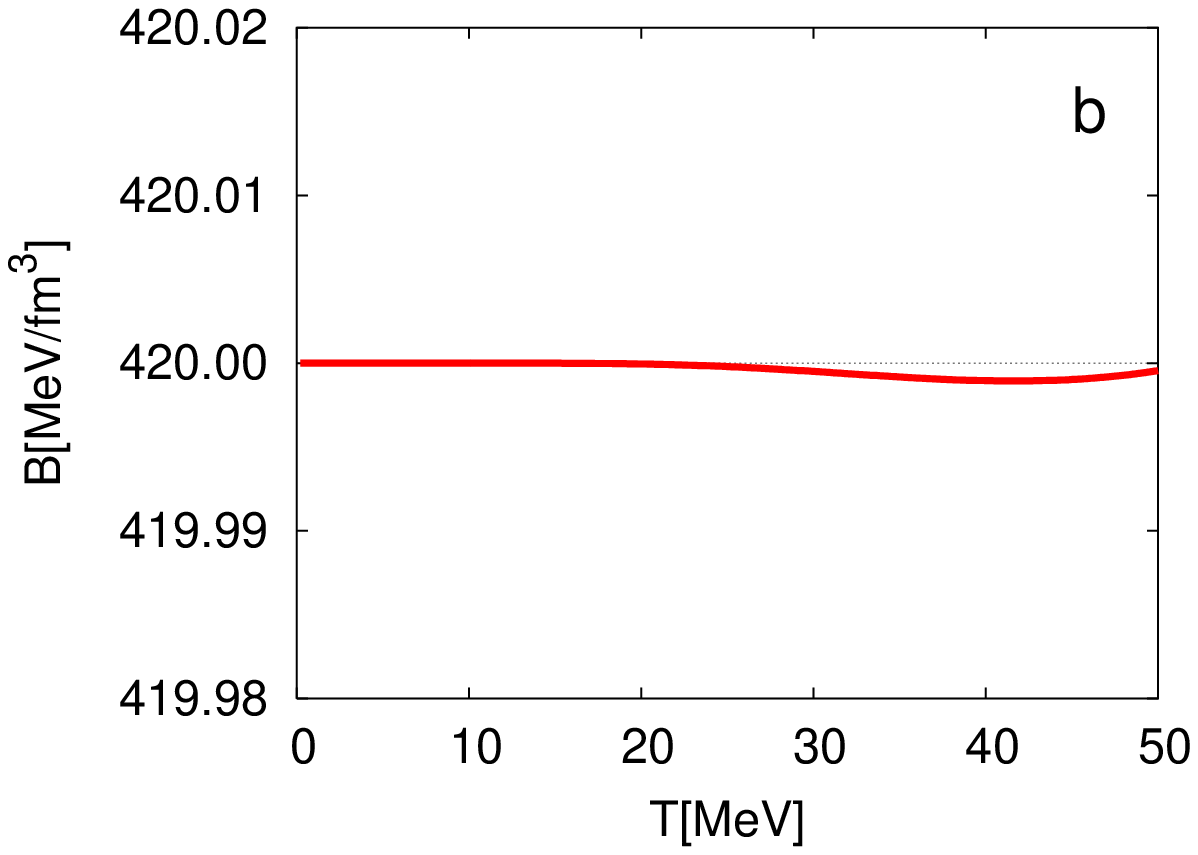}}%
 \caption{\label{fig:1} The  panel (a) depicts ${\mathbf B}$ as a function 
   of $T$ in the baryonic system with the three light quarks. 
   Here ${\cal S}(T)_{q,3}$ is not included in the equation
   of state. The bottom panel gives the same dependence 
   for the baryonic bag of the three strange quarks. }
 \end{figure}
 
 \begin{figure}
\centerline{\includegraphics[width=12.cm]{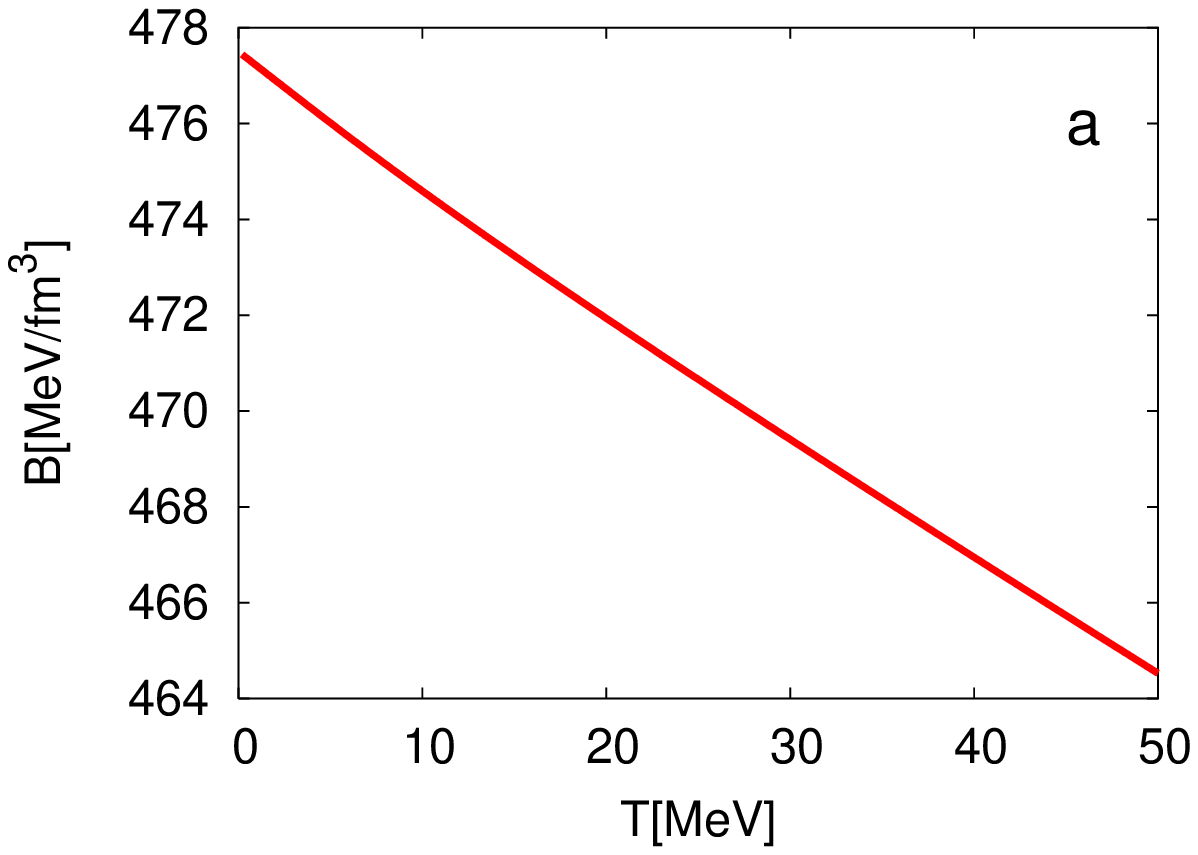}}%
\centerline{\includegraphics[width=12.cm]{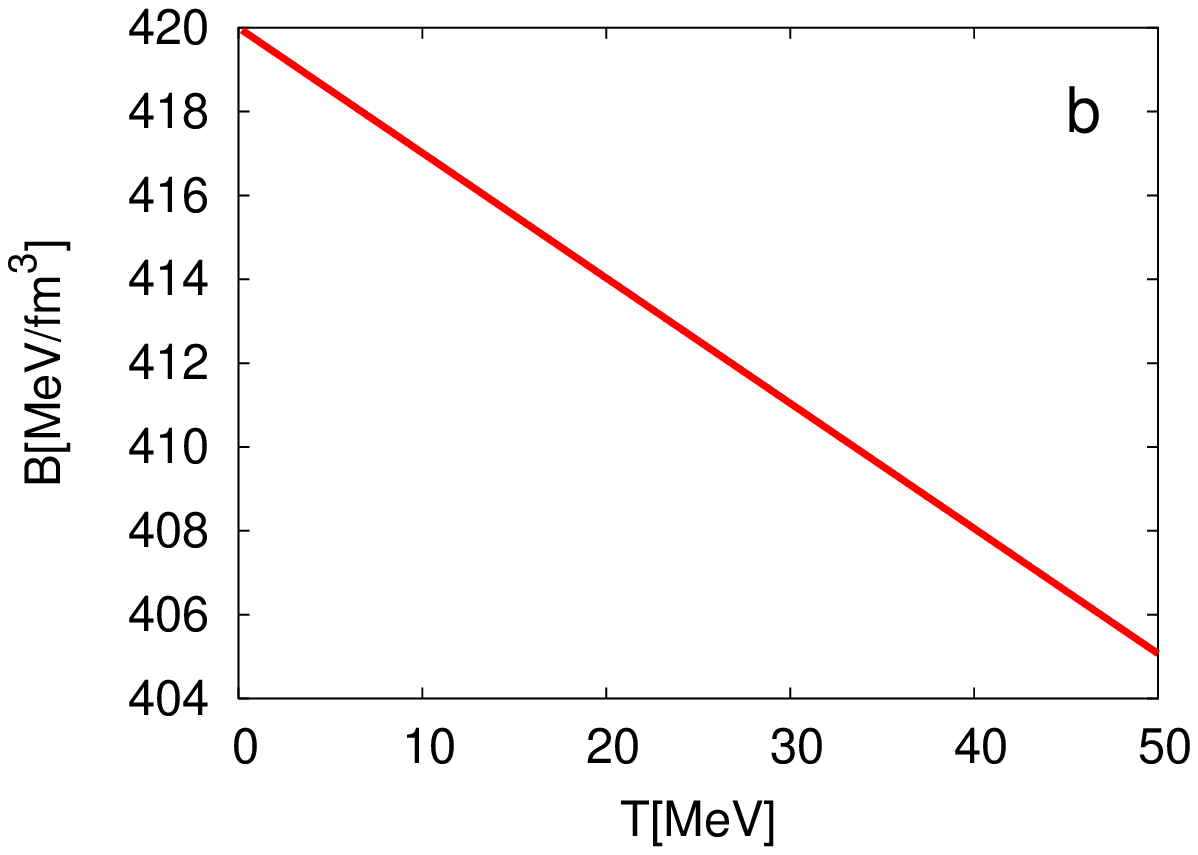}}%
 \caption{\label{fig:2} The top panel shows the thermal structure of
   ${\mathbf B}$ for the three light quarks. Similar for the three
   strange quarks appears in (b). 
   For both of them we added the ground state entropy ${\cal S}(T)_{q,3}$
   in Eq.~\ref{eq:2}.} 
 \end{figure}

In Fig.~\ref{fig:1}(a) we plot ${\mathbf B}$ as a function of $T$
for the baryonic state $\Delta$  while leaving out in Eq.~\ref{eq:2} the
quantum entropy contribution~\cite{Mill,MiTa}. Here we 
have used the vacuum expectation value $1.91\;$GeV/fm$^3$. 
We notice that ${\mathbf B}$ remains constant, especially at very low
temperatures.  Afterwards it only slightly increases so that over
$30\;$MeV temperature, the bag pressure only gains $0.04\;$MeV/fm$^3$. The
reason for the increasing is that the thermal pressure of baryon
constituents (Eq.~\ref{eq:3}) increases
faster than $T s(T)$ (Eq.~\ref{eq:5}).  But at low temperature
($T<20\;$MeV) both pressure and entropy density are too small 
and consequently almost equally badly. Therefore, ${\mathbf B}$
clearly remains constant. Another reason for increasing ${\mathbf B}$ for
$T>20\;$MeV may be the absence of the chemical potential. If the
calculations were done for finite $\mu_q$, we have to modify
equations~\ref{eq:3}~and~\ref{eq:5} and correspondingly add the positive term
$3\mu_q \rho(T,\mu_q)$, where $\rho(T,\mu_q)$ is the quark number
density. In this case $T s(T)$ together with this term may be able to
balance the increasing $p(T)$. Therefore,
we can conclude that ${\mathbf B}$ almost remains constant for a wide range
of temperatures.

     In Fig.~\ref{fig:1}(b) we show the corresponding results for the baryonic
state $\Omega$, again without the quantum entropy contribution. Here we
have used the vacuum  expectation value $1.68\;$GeV/fm$^3$. In this case we
note the range of $T$ in which ${\mathbf B}$  remains absolute constant is
almost the same as in the case of $\Delta$. Beyond this range it decreases
only marginally. Thereafter it starts to 
increase in a region not shown here. The reason for decreasing  ${\mathbf
  B}$ is just the inverse of that in previous case. For strange quark 
mass, which is very much heavier than the masses of light quarks, the
pressure is much smaller than $T s(T)$. But this is expected to be altered
with the temperature. For larger $T$ we could expect 
that ${\mathbf B}$ will increase in a way similar to
the $\Delta$ bag. 

Thus we may conclude that ${\mathbf B}$ is {\it almost}
constant in both baryonic bags, especially at $T<<T_c$, i.e. at temperatures 
at which the quark condensate can be determined from states of low-energy:
the vacuum and the low-lying hadronic resonances, like pions and
kaons~\cite{KRT1,KRT2}. \\ 
In the high temperature region we expect that the values of
\hbox{$\epsilon(T)-3p(T)$} do not vanish, for which the gluonic radiation 
contributions become very dominant. Thus, we expect that at vanishing
temperatures \hbox{$\epsilon-3p$} approaches $4{\mathbf B}$.
Therefrom, we can determine the region of  $T$, in which the bag
pressure remains relatively unchanged. \\

     The inclusion of the quantum entropy density leads to quite different 
effects. This may be seen in Fig.~\ref{fig:2}. Particularly, in $\Omega$
bag there is a linear decrease of ${\mathbf B}$ with 
increasing $T$. In $\Delta$ there is a very small bend around
$T\sim20\;$MeV. Afterwards, the decay gets linear too. Since the asymptotic
value of the quantum entropy density is finite for very high
temperatures~\cite{MiTa}, ${\mathbf B}$ is expected to continuously
decrease with increasing $T$.

\section{\label{sec:4} Conclusion and Outlook}

Some situations where the ground state entropy has essential applications are
shortly discussed in section~\ref{sec:1}. The hadronic matter is expected
to undergo a phase transition into degenerate quarks at low temperatures and
large chemical potentials. These conditions may be present in the interior
of hybrid stars and large planets and could be achieved in the future
heavy-ion experiments. Present investigation can by applied in
understanding the matter under these extreme conditions. Moreover, the physics
of colored quarks plays a prominent role in understanding the dynamics of
color superconductivity.  In a recent work~\cite{MiTa3} we have calculated
the ground state  entropy for a colored two-quark system and compared it
with the entropy arising from the excitations in the BCS model and in the
Bose-Einstein  condensate (BEC) for bosonic
quark-pairs at low temperatures and high quark chemical potentials.
On the other hand, we need to consider the entropy for colored quarks
in order to be able to calculate the consequences of QCD at low
temperatures. \\ 
     Understanding the thermal behavior of colored quark states is very
 useful for different actual applications. It may well bring about a device 
for the further understanding of the physics behind the recent lattice
results for the entropy of static quark-antiquark in heavy quark
potential~\cite{KMTZ1}. The 
entropy difference in a quark-antiquark singlet state on the lattice gives
a value of $2\ln 3$ at vanishing temperature~\cite{Felix2}. Our
estimation for the quantum entropy of {\it one} quark as a part of colorless
singlet state~\cite{Mill} yields the value of $\ln3$~\cite{KMTZ1}. Taking
this in mind, we can also reproduce the lattice entropy results at finite
temperatures~\cite{KMTZ1}. In doing it, we take this $T$-independent term
together with string model and the pure $SU(3)$ gauge theory
results. Furthermore, we believe that the investigation of the quantum
subsystems at finite temperature 
might be useful for understanding the concept of confinement, which could
exists everywhere throughout \hbox{$T\in[0,\infty]$}. The quark distributions
inside the hadron bags reflect themselves as entanglement or - in 
our language - quantum entropy. \\   
~\\
Finally, we can conclude that the presence of the ground state (quantum)
entropy density arising from the $SU(3)_c$ color symmetry in the equation
of state provides a strong temperature dependence for the bag pressure
${\mathbf B}$. The inclusion of this entropy in the equation of state
for the two considered baryonic structures provides their constituents with
an additional heating energy and therefore leads to an almost linear 
decline in ${\mathbf B}$ with increasing $T$. We have contrasted these
results in  both cases to the same properties without the quantum
entropy. Although, there are other possible contributions
to the quantum entropy arising from different physical quantities, like the
spin, however in this model the effects from the color
degrees of freedom are responsible for the decrease in ${\mathbf B}$. \\

     Based on these results we plan further studies about the behavior 
of the structure of confined quark matter at very low temperatures and 
small quark chemical potentials. A further consideration of the idea that
glueballs could appear as $0^{++}$ state in a BEC~\cite{HanJoPet} seems to
be a quite promising further point. The importance of the glueball degrees
of freedom for describing the  hadronic phase for temperature below $T_c$
has already been investigated ~\cite{KRT1,KRT2}. Also the existence of
spin-color waves in the bag we would like to study further. \\ \\

Finally, we wish to acknowledge the stimulating discussions with
Frithjof~Karsch, Krzysztof~Redlich and Helmut~Satz. D.E.M. is very grateful
to the Pennsylvania State University Hazleton for the sabbatical leave of
absence and to the Fakult\"at f\"ur Physik der Universit\"at Bielefeld.


\begin{thebibliography}{10}

\bibitem{Planck64}
M.~Planck,
\newblock {\em Vorlesungen {\"u}ber Thermodynamik},
\newblock Walter De Gruyter \& Co., Berlin, 1964

\bibitem{Schr}
E.~Schr{\"o}dinger,
\newblock {\em Statistical Thermodynamics},
\newblock Cambridge University Press, Cambridge, 1946

\bibitem{Geno}
M.~Genovese,
\newblock {\em About entanglement properties of kaons and tests of hidden
  variables  models}, 
\newblock {arXiv:quant-ph/0305087}

\bibitem{Deutsch}
D.~Deutsch,
\newblock {\em Phys.\ Rev.\ Lett.} 50:631, 1983


\bibitem{Mill}
D.~E. Miller,
\newblock {\em Entropy for $SU(3)_c$ quark states},
\newblock {arXiv:hep-ph/0306302} accepted for publication in Eur.~Phys.~J.~C

\bibitem{MiTa}
D.~E. Miller and A.~Tawfik,
\newblock {\em Entropy for colored quark states at finite temperature},
\newblock {arXiv:hep-ph/0308192}

\bibitem{DoGoHo}
J.~F.~Donoghue, E.~Golovich and B.~R.~Holstein,
\newblock {\em Dynamics of the Standard Model}.
\newblock Cambridge University Press, Cambridge, 1992


\bibitem{Berg:1986aq}
B.~Berg, J.~Engels, E.~Kehl, B.~Waltl, and H.~Satz,
\newblock {\em Z.\ Phys.\ C} 31:167,~1986

\bibitem{SVZ}
M.~A.~Shifman, A.~I.~Vainshtein and V.~I.~Zakharov,
\newblock {\em Nucl.\ Phys.\ B}, 147:385, 1979

\bibitem{ReRuYa}
L.~J. Reinders, H.~Rubenstein and S.~Yazaki,
\newblock {\em Phys.\ Rep.}, 127:1, 1985

\bibitem{Leut}
H.~Leutwyler, 
\newblock {{\em Deconfinement and Chiral Symmetry}, {\bf 2}, 693:716},
\newblock in {\em QCD 20 Years Later}, 
\newblock {eds. P.~M.~Zerwas and H.~A. Kastrup}, 
\newblock (World Scientific, Singapore, 1993)

\bibitem{BoMi}
G.~Boyd and D.~E. Miller,
\newblock {\em The temperature dependence of the $SU(N_c)$ gluon condensate
  from lattice gauge theory},
\newblock {arXiv:hep-ph/9608482}

\bibitem{Milli}
D.~E. Miller,
\newblock {\em Acta\ Phys.\ Polon.\ B}, 28:2937, 1997

\bibitem{Nari}
S.~Narison,
\newblock {\em QCD Spectral Sum Rules}.
\newblock World Scientific, Singapore, 1989

\bibitem{PDG}
K.~Hagiwara, {\it et.~al},
\newblock {\em Review of Particle Physics}.
\newblock {\em {Phys.\ Rev.\ D}},~66:1,~2002

\bibitem{HanJoPet}
T.~H. Hansson, K.~Johnson and C.~Peterson,
\newblock {\em Phys.\ Rev.\ D},~26:2069,~1982

\bibitem{MR}
B.~Muller and J.~Rafelski,
\newblock {\em Phys.\ Lett.\ B}, 101:111 1981 
 
\bibitem{FJL}
R.J.~Furnstahl, X.~Jin and D.~B. Leinweber, 
\newblock {\em Phys.\ Lett.\ B},~387:253,~1996 


\bibitem{KRT1}
F.~Karsch, K.~Redlich and A.~Tawfik,
\newblock {\em Eur.\ Phys.\ J.\ C.}, 29:549, 2003

\bibitem{KRT2}
F.~Karsch, K.~Redlich and A.~Tawfik,
\newblock {\em Phys.\ Lett.\ B}, 571:67, 2003

\bibitem{MiTa3}
D.~E. Miller and A.~Tawfik,
\newblock {\em Entanglement in Condensates involving strong interactions},
\newblock {arXiv:hep-ph/0312368}

\bibitem{KMTZ1}F.~Karsch, D.~Miller, A.~Tawfik and F.~Zantow, in
    progress

\bibitem{Felix2}O.~Kaczmarek, S.~Ejiri, F.~Karsch, E.~Laermann and
        F.~Zantow,  {\it Heavy quark free energies and the renormalized
        Polyakov loop in full QCD}, hep-lat/0312015

\end{thebibliography}
\end{document}